\documentclass[%
aps,
prd,
preprint,
groupedaddress,
nofootinbib,
amsmath,
amssymb]{revtex4-1}

\usepackage{indentfirst}  
\usepackage{bm}    
\usepackage{graphicx}  
\usepackage{latexsym}
\usepackage{graphicx}
\usepackage{psfrag}
\usepackage{epsfig}
\usepackage{amsmath}
\usepackage{tocloft}
\usepackage{array}
\usepackage{setspace}
\usepackage{amssymb}
\usepackage{float}
\usepackage{multirow}
\usepackage{slashbox}
\usepackage{color}
\usepackage[caption=false]{subfig}
\usepackage{epstopdf}
\usepackage[table]{xcolor}

\begin{document}

\title{Very high frequency gravitational waves from magnetars and gamma-ray bursts}   



\author{Hao Wen}
\email[]{wenhao@cqu.edu.cn}
\affiliation{Department of Physics, Chongqing University, Chongqing 401331, P.R. China}

\author{Fangyu Li}
\affiliation{Department of Physics, Chongqing University, Chongqing 401331, P.R. China}

\author{Jin Li}
\affiliation{Department of Physics, Chongqing University, Chongqing 401331, P.R. China}

\author{Zhenyun Fang}%
\affiliation{Department of Physics, Chongqing University, Chongqing 401331, P.R. China}

\author{Andrew Beckwith}%
\affiliation{Department of Physics, Chongqing University, Chongqing 401331, P.R. China}


\date{\today}

\begin{abstract}
\indent Extremely powerful astrophysical electromagnetic (EM) systems could be possible sources of high-frequency gravitational waves (HFGWs).
Here, based on properties of magnetars and gamma-ray bursts (GRBs),
we address ``Gamma-HFGWs'' (with very high-frequency around $10^{20}$~Hz) caused by ultra-strong EM radiation (in the radiation-dominated phase of GRB fireballs)
interacting with super-high  magnetar surface magnetic fields ($\sim10^{11}$~T). By certain parameters of distance and power, the Gamma-HFGWs would have far field energy density $\Omega_{gw}$ around $10^{-6}$, and they would cause perturbed signal EM waves of $\sim10^{-20}$ W/m$^2$ in a proposed HFGW detection system based on the EM response to GWs.
Specially, Gamma-HFGWs would possess distinctive envelopes with characteristic shapes depending on the particular structures of surface magnetic fields of magnetars, which could be exclusive features helpful to distinguish them from background noise. Results   obtained suggest that magnetars could be involved in possible astrophysical EM sources of GWs in the very high-frequency band, and Gamma-HFGWs could be  potential targets for observations in the future.\\


\end{abstract}

\maketitle

\section{Introduction}
\label{sectIntro}
\indent LIGO has announced four direct detections of gravitational waves (GWs), in the intermediate frequency band, from the physical situation of GW sources occurring due to black hole mergers \citep{PhysRevLett.116.061102,secondLIGOGW,PhysRevLett.118.221101,PhysRevLett.119.141101}. This great discovery may inaugurate the era of GW astronomy, and it will also arouse strong interest in looking for GWs from various types of sources in different frequency bands (low, intermediate, high, and very high-frequency bands). In this article, we focus on the possible generation of very high-frequency GWs (around $10^{20}$~Hz) from super-powerful astrophysical electromagnetic (EM) sources. \\
\indent Actually, the generation of GWs from EM sources, as well as the interaction between  GWs  and EM  fields, 
has been studied for a long time. Examples include B-mode polarization in the cosmic microwave
background (CMB) caused by very low-frequency primordial (relic) GWs  \citep{ZhangYang2005CPL,zhaowen2014,Zhao_PRD083006_2006,PhysRevLett.113.021301,Baskaran_PRD083008_2006,
	Polnarev_MNRAS1053_2008,Seljak_PRL2054_1997,Pritchard_AnnPhysNY2_2005},
GWs generated by high-energy astrophysical plasma interacting with intense EM radiation \citep{Servin_PRD68_2003},
GWs produced by EM waves interacting with background magnetic fields \citep{Gertsenshtein_SovPhysJETP_1962,Boccaletti_NuovoCim70_1970,Li.Fang-Yu.120402},
and the EM response to HFGWs  which would lead to perturbed signal EM waves
\citep{LiNPB2016,FYLi_PRD67_2003,FYLi_EPJC_2008,FYLi_PRD80_2009,PRD104025,WenEPJC2014,Shi20065041,Jin2009922,LiXin2016}.
For such issues, the physical conditions and factors of the EM systems, like their strength, structure and scale,
will crucially influence the energy and distribution  of the generated GWs and the way the perturbed signal EM waves appear (e.g., in proposed HFGW detectors).\\
\indent Therefore, some celestial
bodies with extraordinary EM environments, such as magnetars (which have ultra-high surface magnetic fields), would act as natural astrophysical laboratories
to provide extremely strong EM systems as possible GW sources. Thus, in some possible cases, e.g.  a binary system consisting of a magnetar and another celestial body which could emit super-powerful radiation as gamma-ray bursts (GRBs), or a magnetar which emits GRBs itself, the system would be a strong EM source of HFGWs by providing powerful EM waves to interact with  ultra-high magnetic fields. In this paper, we address such ``Gamma-HFGWs'' and their possible characteristic properties.\\

\indent Specifically, the Gamma-HFGWs would be produced by high-energy radiations of GRBs (up to $10^{53}$~erg or even higher \citep{RevModPhys.76.1143})
interacting with the super strong surface magnetic fields of the magnetar ($\sim10^{11}$~T) \citep{Metzger.AstroPhysJourn.659.561.2007}. However, for conservative calculation, we only consider the contribution of such high-energy radiation within the radiation-dominated phase in the fireball of a GRB, where the  energy density  of radiation   decays quite fast, by $r^{-4}$ (because of the conversion of radiation photons
into electron-positron pairs during the transition process into the matter-dominated phase \citep{Piran1999575}).
Based on the Einstein-Maxwell equations \citep{Gertsenshtein_SovPhysJETP_1962,DeLogi_PRD16_1977} in the framework of
general relativity, such radiation and surface
magnetic fields will provide us with a quickly varying energy-momentum tensor $T^{\mu\nu}$ as a powerful HFGW source.
By typical parameters, we estimate that the Gamma-HFGWs (with very high-frequency $\sim10^{20}$~Hz) would have an energy density $\Omega_{gw}$ around $10^{-6}$ 
at an observational distance of $\sim Mpc$ away from the source.
This level of $\Omega_{gw}$ could cause perturbed signal EM waves of strength $\sim10^{-20}$~W/m$^2$ in a  proposed
HFGW detector based on the EM response to
HFGWs and the synchro-resonance effect \citep{FYLi_PRD67_2003,LiNPB2016,FYLi_EPJC_2008,FYLi_PRD80_2009,PRD104025,WenEPJC2014,Li.Fang-Yu.120402}. \\

\indent Only components
of magnetic fields which are perpendicular to
the direction of propagation of GRB radiation will contribute to the generation of  Gamma-HFGWs \citep{Gertsenshtein_SovPhysJETP_1962,DeLogi_PRD16_1977} (this case can be called the ``perpendicular condition''). Thus, the angular distributions of Gamma-HFGWs will appear in specific special patterns (e.g.  the equator-maximum-pattern     or quadrupole-like pattern) according to the specific mode and structure of the surface magnetic fields (their exact structure is still unknown so far, so we here employ a typical possible form \citep{PhysRevD.64.083008} as an example for this paper).    
The misalignment of the rotational axis and
magnetic axis of a magnetar would lead to particular pulse-like envelopes of energy density of Gamma-HFGWs in
the observational direction.
Such unique envelopes would be distinctive properties and criteria to distinguish the signals of Gamma-HFGWs from background noise.\\

\indent This paper is structured as follows.
In Section 2, we present the
form used for the super strong surface magnetic field of the magnetar.
In Section 3, the 
generation of Gamma-HFGWs and their energy density are estimated.
In Section 4, the characteristic envelopes of Gamma-HFGWs are expressed.
In Section 5, we give a summary and conclusions, and present discussion of the consequences of our derivation.

\section{Model of super strong surface magnetic fields of  magnetar}
\label{sectBfields}
The structure of magnetic fields of magnetar, will delineate a magnetar's interaction with EM waves. However, so far, we are not sure about
the concrete form of surface distribution of such magnetic fields, although it's well-known that magnetars can have extremely strong surface magnetic fields reaching to $\sim10^{11}Tesla$
or even higher \citep{Metzger.AstroPhysJourn.659.561.2007}.  Thus, in this paper, we take a typical form \citep{PhysRevD.64.083008} of magnetic fields of a magnetar as an example and basis for calculations in the later sections, i.e., the surface magnetic fields for a magnetar could be generally expressed as \citep{PhysRevD.64.083008}:

\begin{eqnarray}
	\label{eqB}
	&~&\textbf{B}^{surf}=\vec{\bigtriangledown}\times(\vec{r}\times\vec{\bigtriangledown}S),
\end{eqnarray}

Where the symbol $\vec{\bigtriangledown}$ represents the three-dimensional vector differential operator referring to the scale factors $h_i$ to describe the geometry of $t=constant$ spacelike hypersurfaces with the line element different to that in a flat spacetime [see details in the Appendix A of ref. \citep{PhysRevD.64.083008}].   Here we use the spherical coordinates with the orthonormal basis of $\textbf{e}_r$, $\textbf{e}_{\theta}$ and $\textbf{e}_{\phi}$; the $\textbf{r}=r\textbf{e}_r$, and the $\textbf{B}=B_r\textbf{e}_r+B_{\theta}\textbf{e}_{\theta}e+B_{\phi}\textbf{e}_{\phi}$.  The scalar function $S$ can be expanded in a series of spherical harmonics:
\begin{eqnarray}
	\label{eqS}
	S=S(l,m)=S_l^m(r)Y_l^m(\theta,\phi),\nonumber\\
	and~Y_l^m(\theta,\phi)=P_l^m(\cos\theta)e^{im\phi};
\end{eqnarray}
where $P_l^m(\cos\theta)$ is Legendre polynomial. For $l=1,m=0$ (corresponding to
the dipole mode), Eq. (\ref{eqS}) gives:
\begin{eqnarray}
	\label{eqS10}
	S(1,0)&=&C\frac{\cos\theta}{r^2}\sum _{\nu =0}^{\infty }a_{\nu}(\frac{2M}{r})^{\nu},\nonumber\\
	a_0&=&1, a_{\nu}=\frac{(1+\nu)^2-1}{(3+\nu)\nu}a_{\nu-1},(for~\nu\geq1),\nonumber\\
\end{eqnarray}
Define the metric $h$  as \citep{PhysRevD.64.083008}:
\begin{eqnarray}
	\label{eqMandh}
	~h=h(r)=(1-\frac{2M}{r})^{-\frac{1}{2}}, M=\frac{G{m(r)}}{c^2},
\end{eqnarray}

\begin{figure}[!htbp]
	\centerline{\includegraphics[width=7cm]{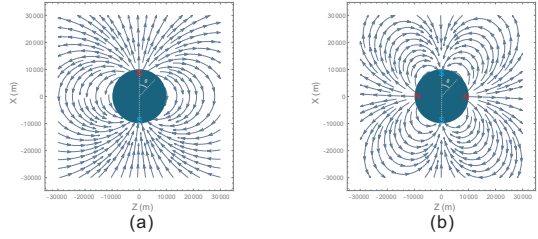}}
	\begin{spacing}{1.2}
		\caption{          Two-dimensional presentations of models of magnetar surface magnetic fields in dipole and quadrupole modes. Poloidal components of dipole mode (a) and quadrupole mode (b)  surface magnetic
			fields of magnetar reach their maximum at polar angle $\theta=\pi/2$ and $\theta=\pi/4,~3\pi/4$ respectively.
			Patterns of surface magnetic fields will crucially influence the angular distributions of Gamma-HFGWs            }
		\label{diandquadrupole2D}
	\end{spacing}
\end{figure}

From Eqs. (\ref{eqB}) to (\ref{eqS10}), magnetic field   (the dipole component)   can be given,
\begin{eqnarray}
	\label{eq_Bsurf}
	&~&\textbf{B}^{surf}(1,0)=\vec{\bigtriangledown}\times( \vec{r}\times\vec{\bigtriangledown}S(1,0)  )\nonumber\\
	&=&C_1\cos\theta\frac{1}{r^3}\sum _{\nu =0}^{\infty }a_{\nu}(\frac{2M}{r})^{\nu}\vec{e}_r\nonumber\\
	&+&C_1\sin\theta\frac{1}{r^3h}\sum _{\nu =0}^{\infty }(\nu+1)a_{\nu}(\frac{2M}{r})^{\nu}\vec{e}_{\theta},
\end{eqnarray}
By calculations of the summation terms in Eq. (\ref{eq_Bsurf}), a typical analytical expression of the surface magnetic field
of the magnetar could be written as [see Fig.\ref{diandquadrupole2D}(a)]:
\begin{eqnarray}
	\label{eq_Bsurf_dipole}
	&~&\textbf{B}^{surf}_{di}(1,0)\nonumber\\
	&=&2C_1\cos\theta\frac{1}{r^3}\frac{-3 r [r^2 \log (1-\frac{2 M}{r})+2 M (M+r)]}{8M^3}\vec{e}_r\nonumber\\
	&+&C_1\frac{\sin\theta}{r^3h}\frac{3 r^2 [2 M (\frac{M}{r-2 M}+1)+r \log (1-\frac{2 M}{r})]}{4M^3}\vec{e}_{\theta},\nonumber\\
\end{eqnarray}

In Eq. (\ref{eqMandh}) we have that $m(r)$ is  the  mass
function that determines the total mass enclosed within the sphere of radius $r$, and \mbox{$m(r)\equiv$~magnetar mass} in our case
because we only concern magnetic fields outside magnetars. Here, $C_1$ and $C_2$ (see below) are constants that have been
calibrated to typical strengths of surface magnetic fields (e.g. $10^{11}T$).\\
\indent Similarly,  for the case of $l=2$, $m=0$, we have
the quadrupole form of surface magnetic fields [Fig.\ref{diandquadrupole2D}(b)]:
\begin{eqnarray}
	\label{eq_Bsurf_quadrupole}
	&~&\textbf{B}^{surf}_{quad}(2,0)
	=3C_2(3\cos^2\theta-1)\frac{1}{r^4}\nonumber\\
	&\cdot&\frac{-3 r [r^2 \log (1-\frac{2 M}{r})+2 M (M+r)]}
	{8M^3}\vec{e}_r\nonumber\\
	&+&3C_2\cos\theta\sin\theta\frac{1}{r^4h}\nonumber\\
	&\cdot&\frac{3 r [2 M (\frac{4 M^2}{r-2 M}+M+r)+r^2 \log (1-\frac{2 M}{r})]}
	{8M^3}\vec{e}_{\theta},\nonumber\\
\end{eqnarray}

For magnetars, the  surface magnetic fields in  quadrupole mode would have comparable
strength to that of  dipole mode \citep{APJ.688.1258}.
In dipole mode,
the poloidal components [see $\vec{e}_{\theta}$ component in Eq. (\ref{eq_Bsurf_dipole})]
have the maximal values at polar angle $\theta=\pi/2$ [Fig.\ref{diandquadrupole2D}(a)],
and the radial components [see $\vec{e}_{r}$ component in Eq. (\ref{eq_Bsurf_dipole})]
have their maximum around polar angle $\theta=0$ and $\pi$ (two magnetic poles).    The $\theta=\pi/2$ means the direction perpendicular to the magnetic axis of a magnetar, and $\theta=0$ means   the direction in the magnetic axis.   
Differently, in quadrupole mode the poloidal components [see $\vec{e}_{\theta}$ component
in Eq. (\ref{eq_Bsurf_quadrupole})] have maximal values around
$\theta=\pi/4$ and $3\pi/4$ [Fig.\ref{diandquadrupole2D}(b)]. These particular distributions of surface magnetic fields
will act key roles to determine the angular distributions of the Gamma-HFGWs generated by EM sources from magnetars (see following sections).

\section{Gamma-HFGWs from magnetars and GRBs}
\label{sectGammaHFGWs}

\indent
It is safe to state that the extremely powerful radiations (around $\sim10^{51}$ to $10^{53}erg$  or even higher in a few seconds) make GRBs
the most luminous (electromagnetically) objects in the Universe \citep{Piran1999575,KulkarniNature1998,KulkarniNature1999}.
According to general relativity, interactions between such radiation of high energy EM bursts and
ultra-intense surface magnetic fields of magnetars ($\sim10^{11}T$
or higher \citep{Metzger.AstroPhysJourn.659.561.2007}), can
provide a fast varying energy momentum tensor $T^{\mu\nu}$ as a strong EM source of
HFGWs in very high-frequency band (denoted as ``Gamma-HFGWs'', the same hereafter). \\
\indent Lots of models
of inner engine of GRBs had been proposed to explain the origin of so huge amount of energy, such as black-hole
accretion, collapsar model, supernova model (see review by Piran \citep{RevModPhys.76.1143}),
binary neutron star mergers \citep{EichlerNature1989, NarayanAPJ1992},
black hole-neutron star mergers \citep{Paczynski1991},
Blandford-Znajek mechanism \citep{Blandford01071977},  pulsar model \citep{RevModPhys.76.1143,Uso1992,Usov15041994,Smolsky1996,0004-637X-531-2-764,Drenkhahn2002,Spruit2001},
magnetar model \citep{ThompsonDuncan2001,Eichler01102002,1538-4357-552-1-L35,ThompsonArxiv2006}, etc.
Specially, the magnetar model of GRBs with fireball scenarios are studied by some previous works \citep{0004-637X-702-2-1171, 0004-637X-606-2-1000, refId0, 0004-637X-791-1-46, 1009-9271-6-S1-44, refId2}, and a
possible case is to consider fireball
trapped near the magnetar surface by the super strong magnetic fields \citep{Thompson15071995, 0004-637X-561-2-980,
	0004-637X-710-2-1335,AIP2943498, 0004-637X-558-1-237, 0004-637X-677-1-488, 0004-637X-685-2-1114}. Therefore, no matter in the case that GRBs source would combine magnetar as a binary system, or in the case that magnetar itself would become the source of GRBs, once such  GRB radiations interact with the magnetar surface magnetic fields, it could lead to considerable generation of Gamma-HFGWs.\\

\indent GRBs have complicated process and mechanism, especially for the problem of inner engine that produces the relativistic energy flow \citep{Piran1999575}.
According to fireball internal-external shocks model, the generation of observed GRBs would be on account of the process of kinetic energy of
ultra-relativistic flow to dissipate during the internal collisions (internal shocks) \citep{RevModPhys.76.1143}.
Piran had summed \citep{Piran1999575} generic pictures to suggest that in the fireball model the GRBs are composed of several stages:
(i) a compact inner ``engine'' to produce
a relativistic energy flow, (ii) stage of energy transportation,
(iii) conversion of this energy to observed prompt radiation, (iv) conversion of the remaining energy to afterglow.\\
\indent For stage (i), Goodman \citep{Goodman1986} and Paczynski \citep{Paczynski1986} proposed the relativistic fireball model and had shown that the sudden release
of a large quantity of gamma-ray photons into such compact region  can lead to an opaque
photon-lepton fireball (pairs-radiation plasma, by production of electron-positron pairs from photon-photon scattering) \citep{Piran1999575},
because if the photon energy reaches high enough ($>511KeV$), electron-positron pairs can  be formed from the radiations.  \\

\begin{figure}[!htbp]
	\centerline{\includegraphics[width=8cm]{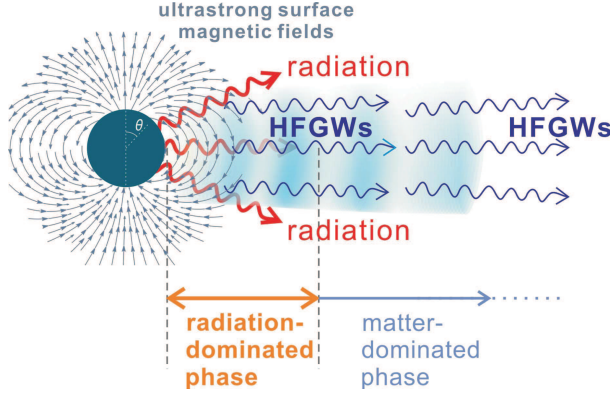}}
	\begin{spacing}{1.2}
		\caption{  Gamma-HFGWs caused by strong radiations (in radiation-dominated phase of fireball in GRBs) interacting with ultrastrong surface magnetic fields (dipole mode) of magnetar. Expanding fireball has two basic phases:
			radiation-dominated phase (typically $<10^7m$) and  matter-dominated phase ($>10^7m$); in radiation-dominated stage, energy of strong radiation decays quite quickly by $distance^{-4}$ \citep{Piran1999575}.
			However, in very early stage of the radiation-dominated phase, photon energy still dominates in the fireball and can provide extremely powerful EM source to interact with the
			strong surface magnetic fields, i.e. provide a fast varying energy momentum tensor $T^{\mu\nu}$ as a strong EM source of
			generation of Gamma-HFGWs, and such HFGWs are transparent to the optically-thick fireball. This figure is an intuitive demonstration and the scale is not exact.                    }
		\label{fireball}
	\end{spacing}
\end{figure}

\indent Thereafter, pairs-radiation plasma behaves like a perfect fluid and expands by its own
pressure \citep{Piran1999575}. During this expansion and energy transportation stage,
the expanding fireball has two basic phases \citep{Piran1999575}:
a radiation-dominated phase and a matter-dominated phase (see Fig.\ref{fireball}). In early stage of radiation-dominated phase, most  energy comes out as high-energy radiation \citep{Piran15081993},
and the fluid of plasma accelerates in the process of expansion with very
large Lorentz factors, and then a transition from
the radiation-dominated phase to the matter-dominated phase takes place when the fireball has a size about
$10^7m$ (typical value) \citep{Piran1999575}.
Crucially, in the radiation-dominated
stage, energy of radiation decays by $distance^{-4}$ (much faster than normal spherical radiation in free space
which decays  by $distance^{-2}$, due to the formation of electron-positron pairs from photons, see details in sec.6.3 of ref. \citep{Piran1999575}).\\
\indent Therefore, although GRBs have overall extremely complex evolutionary histories, in the early stage of the radiation-dominated phase,
the photon energy still dominates in the fireball and  we have that photon energy can interact with
extremely strong magnetar surface magnetic fields in order to become a considerable EM source of Gamma-HFGWs (Fig.\ref{fireball}).
While, for conservative estimation procedures used here we
will consider only a much shorter interaction range for calculation (of generation of Gamma-HFGWs) as occurring in the very early stage of the radiation-dominate phase, i.e., we are only considering the interaction range from $10^4m$ (supposed magnetar
radius) to $2\times10^4m$.\\

\indent In local area we have that, for a specific propagation direction (i.e. z-direction), the energy flux density of this strong EM source of Gamma-HFGWs can be represented by the
\mbox{``0-3''} component of an  energy momentum tensor $\mathop{T^{03}}\limits^{(1)}$ \citep{landau1975classical, FYLi_EPJC_2008}:
\begin{eqnarray}
	\label{T03}
	&~&\mathop{T^{03}}\limits^{(1)}=\frac{-1}{\mu_0}(F^{0(0)}_{~\alpha}\tilde{F}^{3\alpha(1)}+\tilde{F}^{0(1)}_{~\alpha}{F}^{3\alpha(0)})
	\nonumber\\
	&=&\frac{-1}{\mu_0}[0+\tilde{F}^{0(1)}_{~1}F^{31(0)}]\nonumber\\
	&=&\frac{1}{\mu_0c}\tilde{E}^{(1)burst}_xe^{i(kz-\omega t)}B^{(0)surf}_
	{\theta},
\end{eqnarray}

Here, ``$k$'' and $\tilde{F}^{\mu\nu(1)}$ are wave vector and EM tensor of EM waves of GRBs;
$F^{\mu\nu(0)}$ is an EM tensor of the surface magnetic fields. Here we define the outward radial direction as the z-direction, i.e.,
$F^{31(0)}=-F^{13(0)}=-B^{(0)surf}_y=B^{(0)surf}_{\theta}=10^{11}T$ (simply treat other components as zero, because
only the $B^{(0)surf}_{\theta}$ [poloidal component of the surface magnetic fields. I.e. the  $\vec{e}_{\theta}$
components in Eqs. (\ref{eq_Bsurf_dipole})
and (\ref{eq_Bsurf_quadrupole})] which is perpendicular to the direction of GRBs in supposed given configuration here,
will contribute to the generation of  HFGWs \citep{Gertsenshtein_SovPhysJETP_1962,DeLogi_PRD16_1977} (noted as ``perpendicular condition'', the same hereafter).
Thus, through magnetar surface magnetic fields in dipole-mode,
generated HFGWs will follow  an equator-maximum-pattern, i.e., their angular distribution
mainly concentrate around the region of polar angle $\theta=\pi/2$ (equator area).
What is noticeable is that magnetar surface magnetic fields in
quadrupole-mode, generate Gamma-HFGWs which would also radiate in a quadrupole pattern concentrating around
$\theta=\pi/4$ and $3\pi/4$.\\
\indent For estimating such Gamma-HFGWs, we
can first focus on a very thin layer in local area, with the assumption 
that the radiation and surface magnetic fields
can be treated as being uniform. So the generation of Gamma-HFGWs can be expected to be given by the linearized Einstein field equation as follows:
\begin{eqnarray}
	\label{EinsteinEQ}
	&~&\Box \tilde{h}^{\mu\nu}(z,t)=-\frac{16\pi G}{c^4}T^{\mu\nu}\nonumber\\
	&=&-\frac{16\pi G}{c^4}\cdot\frac{1}{\mu_0c}\tilde{E}^{(1)burst}_x e^{i(kz-\omega_{\gamma} t)} B^{(0)surf}_{\theta},
\end{eqnarray}
A solution of the above linearized Einstein equation can be obtained to be presented as:
\begin{eqnarray}
	\label{Hsolution}
	&~&\tilde{h}^{\mu\nu}(z,t)=A^{\gamma}e^{i(kz-\omega t+\frac{3\pi}{2})}\nonumber\\
	&=&\frac{-z8\pi G\tilde{E}^{(1)burst}_xB^{(0)surf}_{\theta}}{kc^5\mu_0}e^{i(kz-\omega_{\gamma} t+\frac{3\pi}{2})},
\end{eqnarray}
This local solution composed of planar GWs caused by a uniform EM source clearly shows that the accumulation effect (because
the HFGWs caused by radiation will be accumulated during their propagation along
with the radiation synchronously due to their identical speed of light) is proportion to the accumulative distance (term ``$z$''), which is in total accordance to our previous results derived by use of the accumulation effect which is the case of
what happens when we use planar GWs \citep{Boccaletti_NuovoCim70_1970}.\\
\indent However, for our case, using the Gamma-HFGWs, the situation is much more complicated. The background magnetic fields (surface
magnetic fields of the magnetar) will nonlinearly decrease along the radial direction [Eq. (\ref{eq_Bsurf_dipole})
and Eq. (\ref{eq_Bsurf_quadrupole})],
and the radiation will also decay in the ratio of  $\sim1/z^4$ in the radiation-dominated
phase (within distance $<10^7m$) \citep{Piran1999575}. Thus we find that the compositive contribution of processes due to the generation of Gamma-HFGWs is very different to what happens in the scenario of GW generation \citep{Gertsenshtein_SovPhysJETP_1962}  due to the interaction effects of planar EM waves with uniform magnetic fields. E.g., for a certain
power produced by GRB inner engine (noted as $P_{total}^{\gamma}$), we find that the
energy flux density of the EM waves at distance of $r_0$ (radius of magnetar) should be
$P_{total}^{\gamma}/4\pi r_0^2\sim(\tilde{E}^{(1)burst}_{x.r_0})^2/(\mu_0c)$.  Thus, at distance of $r$,
electric component of the radiation is $\tilde{E}^{(1)burst}_x=\tilde{E}^{(1)burst}_{x.r_0}\cdot r_0^2/r^2$. \\
\indent Therefore,  in order to obtain expression of accumulated amplitude of the Gamma-HFGWs ($A^{\gamma}_{accum}$), we can integrate the 
Gamma-HFGWs generated within very thin local layers   [at distance of ``r'', with thickness of $dr$, so we can employ the result of Eq. (\ref{Hsolution})]  from the magnetar surface to a certain
larger distance ``$z$''. If we employ the dipole surface magnetic field [$B^{(0)surf}_{\theta}$, from the
second part of Eq. (\ref{eq_Bsurf_dipole}), i.e. only take the $\vec{e}_{\theta}$ component, because the $\vec{e}_r$ component does not contribute], it can be worked out to read as:
\begin{eqnarray}
	\label{fireballAaccumStrictBsurf}
	&~&A^{\gamma}_{accum}(z)=\int^z_{r_0}\frac{8 \pi  G}{{kc}^5 {\mu_0}}(\tilde{E}^{(1)burst}_{x.r_0}\cdot \frac{r_0^2}{r^2})B^{(0)surf}_{\theta}\cdot
	\frac{r}{z}dr \nonumber\\
	&=&\frac{ 3 C_1 r_0^2\tilde{E}^{(1)burst}_{x.r_0}\frac{8 \pi  G}{{kc}^5 {\mu_0}} }{4 h M^3z}\sin\theta \nonumber\\
	&\cdot&\int^z_{r_0} \frac{r^2 [2 M \left(\frac{M}{r-2 M}+1\right)+r ln\left(1-\frac{2 M}{r}\right)]}{r^4} \, dr \nonumber\\
	&=&\frac{3C_1r_0^2}{z8M^3h}\cdot\frac{8 \pi  G}{{kc}^5 {\mu_0}}\tilde{E}^{(1)burst}_{x.r_0}\sin\theta\nonumber\\
	&\cdot&[-\frac{2M}{z}+ln(1-\frac{2M}{z})+2\text{Li}_2\frac{2M}{z}\nonumber\\
	&+&ln\frac{r_0}{r_0-2M}-2\text{Li}_2\frac{2M}{r_0}+\frac{2M}{r_0} ];
\end{eqnarray}
here, $Li_2(\frac{2 M}{r_0})=\sum_{k=1}^{\infty }\frac{({2 M}/{r_0})^k}{k^2}$ is  polylogarithm function
of  order 2 with argument $\frac{2 M}{r_0}$ (similarly hereafter). 
The amplitude of GW from any layer at distance of $r$, will decay into level $\propto\cfrac{r}{z}$ (in the ratio of a spherical wave) once the GW propagates to the concerned distance $z$, and this is why we have the term ``$\cfrac{r}{z}$'' to the left of ``$dr$'' in the first line of Eq. (\ref{fireballAaccumStrictBsurf}).
The Eq. (\ref{fireballAaccumStrictBsurf}) looks complicated, but actually if we take only the first order of $Li_2(\frac{2 M}{z})$ (i.e., $\frac{2 M}{z}$), so that it has a simple asymptotic behavior  in large distance:\\
\begin{eqnarray}
	\label{fireballAaccumStrictBsurfaymptotic}
	A^{\gamma}_{accum}(z)\rightarrow p1\cdot  z^{-1}+p2\cdot z^{-2}
\end{eqnarray}
where
\begin{eqnarray}	
	p1=\frac{3C_1r_0^2}{M^3h}\cdot\frac{ \pi  G}{{kc}^5 {\mu_0}}\tilde{E}^{(1)burst}_{x.r_0}\sin\theta\nonumber\\
	\cdot(ln\frac{r_0}{r_0-2M}-2\text{Li}_2\frac{2M}{r_0}+\frac{2M}{r_0}) \nonumber.\\
	p2=\frac{3C_1r_0^2}{4M^2h}\cdot\frac{8 \pi  G}{{kc}^5 {\mu_0}}\tilde{E}^{(1)burst}_{x.r_0}\sin\theta;
\end{eqnarray}
\indent Similarly, for  when we derive the surface magnetic fields in a quadrupole mode [$B^{(0)surf}_{\theta-quad}$, from the
second part of Eq. (\ref{eq_Bsurf_quadrupole})], we find that the accumulated amplitude of
Gamma-HFGWs   can be given as:
\begin{eqnarray}
	\label{fireballAaccumStrictBsurfQuad}
	&~&A^{\gamma-quad}_{accum}(z)=\int^z_{r_0}\frac{8 \pi  G}{{kc}^5 {\mu_0}}(\tilde{E}^{(1)burst}_{x.r_0}\cdot \frac{r_0^2}{r^2})B^{(0)surf}_{\theta-quad}\cdot
	\frac{r}{z}dr \nonumber\\
	&=&\frac{9\pi G{C_2}\sin\theta\cos\theta \tilde{E}^{(1)burst}_{x.r_0} r_0^2 }{kc^5\mu_0hM^3z}
	\{ \frac{-2}{r_0}+\frac{2M^2}{3}(\frac{-1}{r_0^3}+\frac{1}{z^3})\nonumber\\
	&+&\frac{2}{z}+\frac{ln(1-2M/r_0)}{r_0} -\frac{ln(1-2M/z)}{z}\nonumber\\
	&+&[ln\frac{r_0}{r_0-2M}+ln(1-\frac{2M}{z})]/M \}.
\end{eqnarray}
\indent So far, it appears    that   lots of confirmed magnetars are in the Milky Way and can have short distance values of $\sim kpc$, but all currently observed GRBs are from distant galaxies outside the Milky Way, and that the nearest one is GRB 980425 with a redshift $z=0.0085$ or about 36 Mpc away. Even if any GRB happens within a distance of $\sim kpc$, it would cause globally ozone depletion and it might lead to great ecological damage and extinction of life on Earth (this had been believed as a possible reason of the late Ordovician mass extinction \citep{GRBextinction}). Therefore, as presented in Table \ref{gammaburstSignal}, if some magnetars in proper distance with GRBs of suitable power, they still could provide far field effect of Gamma-HFGWs on the Earth (or far field observation points) and meanwhile the power of Gamma-ray can decay into a safe level.\\ \indent E.g., if the maximum of surface magnetic fields of magnetar $\sim10^{11}Tesla$, $P_{total}^{\gamma}\sim10^{54}erg\cdot s^{-1}$, magnetar distance $\sim 1~Mpc$, then the energy density $\Omega_{gw}$ of Gamma-HFGW at the Earth could be $\sim 10^{-6}$ [here, $\Omega_{gw}=\frac{\pi^2}{3}h^2(\nu/\nu_H)^2$, where $\nu_H$ is present Hubble frequency, we have that the value of $h$ is given to be the GW amplitude]. Meanwhile, in this case, the power of GRB around the globe is only about $30~Watt\cdot m^{-2}$ which is in a safe level far less than the order of magnitude to cause  global ozone depletion \citep{GRBextinction}. Other possible cases with suitable parameters of distance and GRB power are also shown in the shaded cells in Table \ref{gammaburstSignal}. We can find that, $\Omega_{gw}$ in cells with larger distance than these shaded cells, will have too low energy density for potential detection (e.g. in proposed HFGW detectors \citep{FYLi_PRD67_2003,FYLi_EPJC_2008,LiNPB2016,FYLi_PRD80_2009,Li.Fang-Yu.120402}), and cells with shorter distance than these shaded cells can have higher $\Omega_{gw}$ but will lead to stronger GRB power which would be dangerous to life and existing ecological systems on Earth. Therefore, the shaded cells in Table \ref{gammaburstSignal} present optimal range of Gamma-HFGW sources with proper distance and suitable power (in safe level near globe) to be potentially observational targets of HFGWs from the Earth. Nevertheless, for other cases which cannot provide sizable far field effect on the Earth, such as more faraway GRBs, the possibility still would not be excluded that in the future some spacecraft-based HFGW detector approaching closer area to such sources or some Earth-based detector with greatly enhanced sensitivity would also might be able to capture these Gamma-HFGWs.  \\
\indent Some proposed HFGW detection system \citep{FYLi_PRD67_2003,FYLi_EPJC_2008,LiNPB2016,FYLi_PRD80_2009,Li.Fang-Yu.120402} is especially sensitive to GWs in very high-frequency bands.   E.g., the Gamma-HFGWs ($\Omega_{gw}\sim10^{-6}$) would generate the first-order perturbed signal EM waves having power of $\sim10^{-20}Watt$ per $m^2$ in such planned detection system.  However, issues about how to experimentally extract and distinguish such perturbed EM signals and relevant techniques, are not key points in this paper, and related topics will be addressed in other works. \\

\begin{table*}[!htbp]
	\caption{\label{gammaburstSignal}   For some parameters in possible range (not limit to that have already been confirmed by current observations), we estimate accumulated energy density ($\Omega_{gw}$, at far observational distance) of the Gamma-HFGWs generated by interaction
		between powerful radiation of GRBs and ultra-high surface magnetic fields of magnetars. Representative frequency of Gamma-ray
		(also of the Gamma-HFGWs) is set to $\sim10^{20}Hz$. Shaded cells in the table indicate the parameter range where the Gamma-HFGWs would cause perturbed signal EM waves of $\sim10^{-20}W/m^2$ in proposed HFGW detector  meanwhile corresponding power of GRBs would decay into about $30~Watt/m^2$ (which is safe, since it is far less than the power able to cause globally ozone depletion and to result in extinction of lives). Therefore, the shaded cells represent the Gamma-HFGW sources with optimal parameters to be possible potential observational targets in the future.             }
	\begin{center} \centerline{
	    \begin{tabular*}{110mm}{@{\extracolsep{\fill}}ccccc}
	    	\hline
	    	\hline
	    	Observational & \multicolumn{4}{c}{Dipole mode, $B^{surf}=4\times10^{11}T$,} \\
	    	distance away  & \multicolumn{4}{c}{$P_{total}^{\gamma}$ ($erg\cdot s^{-1}$)} \\
	    	from magnetar &&$3\times10^{54}$&$3\times10^{52}$&$3\times10^{50}$\\
	    	\hline
	    	~& \\
	    	{$\sim ~10kpc$} &$\Omega_{gw}^{\gamma}:$&~$1.2\times10^{-2}$~~&~$1.2\times10^{-4}$~&\cellcolor{gray!25}~$1.2\times10^{-6}$~~\\		
	    	~& \\
	    	{$\sim ~100kpc$} &$\Omega_{gw}^{\gamma}:$~&~$1.2\times10^{-4}$~~&\cellcolor{gray!25}~$1.2\times10^{-6}$~&~$1.2\times10^{-8}$~~\\
	    	~& \\
	    	{$\sim ~Mpc$} &$\Omega_{gw}^{\gamma}:$&\cellcolor{gray!25}$1.2\times10^{-6}$~~&~$1.2\times10^{-8}$~&~$1.2\times10^{-10}$~~\\
	    	~& \\
	    	{$\sim ~10Mpc$} &$\Omega_{gw}^{\gamma}:$&$1.2\times10^{-8}$~~&~$1.2\times10^{-10}$~&~$1.2\times10^{-12}$~~\\			
	    	~& \\
	    	{$\sim ~Gpc$} &$\Omega_{gw}^{\gamma}:$&$1.2\times10^{-12}$~~&~$1.2\times10^{-14}$~&~$1.2\times10^{-16}$~~\\
	    	\hline&
	    \end{tabular*}
	}
	\end{center}
\end{table*}

\section{Characteristic envelopes of Gamma-HFGWs}
\label{envelope}

\indent  Special geometrical information of structure of magnetar surface magnetic fields, could lead to the existence of characteristic envelopes of energy density of Gamma-HFGWs at specific observation directions, and each special feature of
these GW signals can be very helpful  in order to distinguish them from background noise signals.

\begin{figure}[!htbp]
	\centerline{\includegraphics[width=15cm]{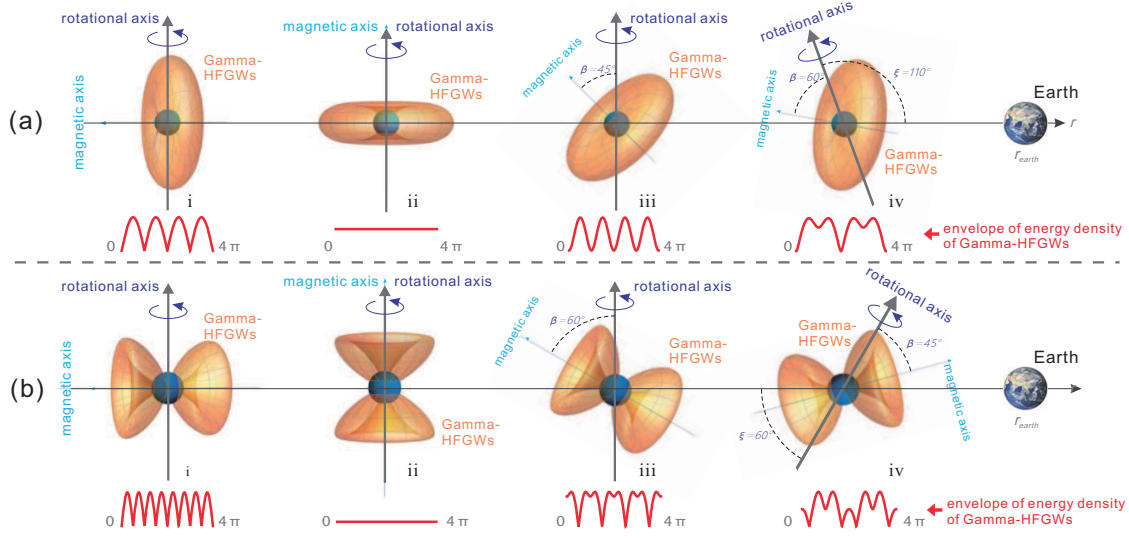}}
	\begin{spacing}{1.2}
		\caption{        Different angular configurations would cause characteristic envelopes of Gamma-HFGWs. This diagram intuitively explains how various distinctive Gamma-HFGW envelopes received at the Earth (or at far field observation point) can be formed according to different angular sets (angle  $\beta$ between rotational  and magnetic axis, and angle $\xi$ between rotational axis and observational direction).
			The energy of Gamma-HFGWs facing 
			the Earth will fluctuate with respect to the rotational phase, and then lead to diverse  envelopes of the energy density of Gamma-HFGWs, similar to the formation of pulsing signals from pulsars.  Sub-figures (a) and (b) show examples of distinctive envelopes by different $\beta$
			and $\xi$, of  equator-maximum pattern  and quadrupole pattern Gamma-HFGWs (see Eqs. \ref{gammappfenvelopeBipole}   and \ref{gammappfenvelopeQuadrupole}).
			Here $\xi$ is $90^{\circ}$ in cases i-iii, and $\beta$ is $90^{\circ}$, $0^{\circ}$, $45^{\circ}$
			and $60^{\circ}$ for cases i to iv, respectively.
			For some angular sets, the envelopes could be more complicated, e.g. in case iv they
			appear in unusual distinctive shapes containing both higher and lower mixed peaks.             }
		\label{compareWaveEnvelopes}
	\end{spacing}
\end{figure}

However, the exact structure of magnetar surface magnetic fields still is unclear so far. Nevertheless, here,  as mentioned above, we can take the form of magnetar magnetic fields \citep{PhysRevD.64.083008} as an example, to present how particular surface magnetic fields lead to corresponding special GW envelopes. In detailed analysis
due to using the fact that the rotational axis and magnetic axis of a given magnetar are usually not identical, we find that during one period, that the maximums of Gamma-HFGWs will not always directly point to the observation direction. Therefore, the envelopes of energy density of these Gamma-HFGWs
will thereby vary and fluctuate periodically according to the rotation of magnetar. This phenomenon is similar to the mechanism of what is seen during the analysis of pulsing signals from pulsars (where the misalignment between these two axis of pulsars usually causes the
peak of EM radiations facing to the Earth once for every spin period).

\begin{figure}[!htbp]
	\centerline{\includegraphics[width=15cm]{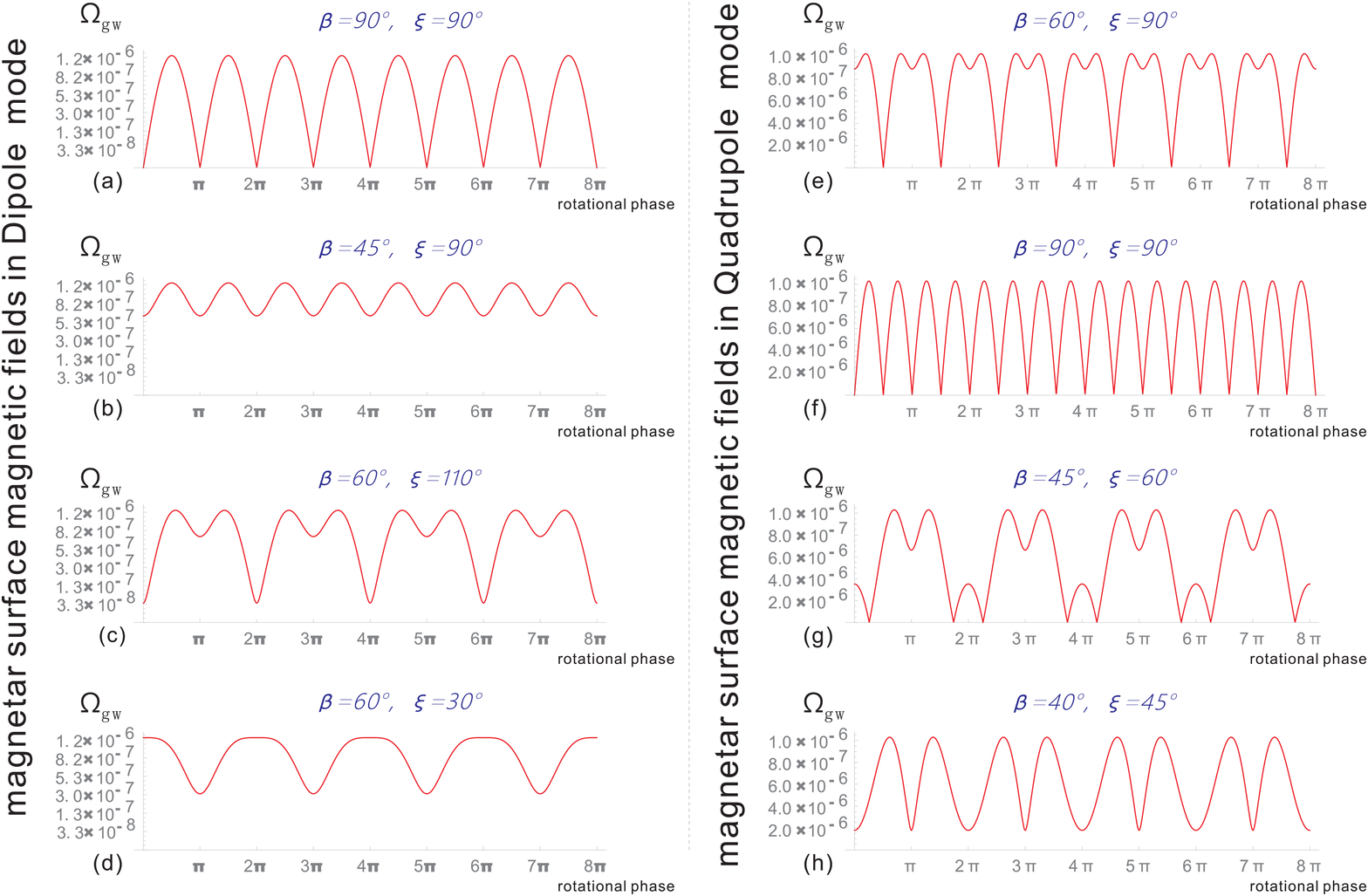}}
	\begin{spacing}{1.2}
		\caption{  Examples of energy density envelopes of Gamma-HFGWs at the Earth or far field observation points. Typical envelopes of energy density of Gamma-HFGW s are estimated for both
			equator-maximum   case (sub-figure (a) to (d), see Eq. \ref{gammappfenvelopeBipole})
			and quadrupole case (sub-figure (e) to (h), see Eq. \ref{gammappfenvelopeQuadrupole}) at various angles $\beta$ and
			$\xi$ (the same  as those defined in Fig. \ref{compareWaveEnvelopes}).
			Parameters for  the equator-maximum   case are: observation distance $=1 Mpc$, $P_{total}^{\gamma}=3\times10^{54}$ erg/s, $Bsurf=4\times10^{11}T$. Parameters for the quadrupole case are:
			$P_{total}^{\gamma}=6\times10^{54}$ erg/s, $B^{surf}=5\times10^{11}T$. These particular envelopes would be helpful to distinguish the
			Gamma-HFGWs from background noise in possible detection schemes.               }
		\label{hfgwGammaSignalweak}
	\end{spacing}
\end{figure}

In Fig.\ref{compareWaveEnvelopes}, we present that in different values of angles (i.e. angle between
magnetic and rotational axis, noted as $\beta$, and angle between rotational axis and observation direction, noted as $\xi$),
envelopes of energy density of Gamma-HFGWs would particularly appear in some pulse-like patterns
with various distinctive shapes.
By using $\Omega_{gw}=\frac{\pi^2}{3}h^2(\nu/\nu_H)^2$ and  if we evaluate Eqs. (\ref{fireballAaccumStrictBsurf}) to (\ref{fireballAaccumStrictBsurfQuad}),
with coordinate transformations, we can have resulting analytical expressions of envelopes of energy density of Gamma-HFGWs  at the Earth or far field observation points (here $\Omega_{gw}^{\gamma-eq}$ and $\Omega_{gw}^{\gamma-quad}$ are for  equator-maximum
and qudrupole cases respectively):
\begin{eqnarray}
	\label{gammappfenvelopeBipole}
	&~&\Omega_{gw}^{\gamma-eq}=\frac{\pi^2}{3}(A^{\gamma}_{accum}|_{\theta=\pi/2})^2(\frac{\nu}{\nu_H})^2\nonumber\\
	&\cdot&[(\sin\xi \cos\beta \cos\varphi+\cos\xi
	\sin\beta)^2+\sin^2\xi  \sin^2\varphi],\nonumber\\
\end{eqnarray}
and for the quadrupole case,
\begin{eqnarray}
	\label{gammappfenvelopeQuadrupole}
	&~&\Omega_{gw}^{\gamma-quad}=\frac{\pi^2}{3}(A^{\gamma-quad}_{accum}|_{\theta=\pi/4})^2(\frac{\nu}{\nu_H})^2\nonumber\\
	&\cdot&(\cos\xi\cos{\beta}-\sin\xi\sin{\beta} \cos\varphi)^2 \nonumber\\
	&\cdot&[(\sin\xi \cos{\beta} \cos\varphi
	+\cos\xi \sin{\beta})^2+\sin^2\xi \sin^2\varphi].\nonumber\\
\end{eqnarray}

\begin{table*}[!htbp]
	\caption{\label{effectiveGammaHFGWsTable}    Given different effective accumulation distance D of Gamma-HFGWs sources, decay parameter $\Lambda$ of effective radiation in GRBs, decay parameter $\Xi$ of
		effective strong background magnetic fields, estimated energy density  of effective-Gamma-HFGWs ($\Omega^{\gamma-eff}_{gw}$) in far field regions are given.
		Here $P_{total}^{\gamma}$ is assumed $5\times 10^{54} erg/s$, and far field observational distance sets to $3.3Mpc$ from the magnetar.         }
	\begin{center} 
	\begin{tabular}{ccccccc }
		\hline
		\hline
		effective accumulation& & \multicolumn{5}{c} {far field (3.3Mpc) $\Omega^{\gamma-eff}_{gw}$  by different decaying parameters $\Lambda$ and $\Xi$} \\
		distance around source&&$\Lambda=2$, $\Xi=1$;&$\Lambda=2$, $\Xi=2$;&$\Lambda=2$, $\Xi=3$;&$\Lambda=2$, $\Xi=4$;&$\Lambda=4$, $\Xi=5$ \\
		\hline
		{$D=10~km$} &$\Omega^{\gamma-eff}_{gw}:$&$5.3\times10^{-6}$~~&~$3.0\times10^{-6}$~&~$1.8\times10^{-6}$~~&~$1.1\times10^{-6}$~~&~$4.3\times10^{-7}$~ \\
		~\\
		{$D=100~km$} &$\Omega^{\gamma-eff}_{gw}:$&$1.7\times10^{-5}$~~&~$5.2\times10^{-6}$~&~$2.3\times10^{-6}$~~&~$1.3\times10^{-6}$~~&~$4.3\times10^{-7}$~ \\
		\hline
	\end{tabular}
	\end{center}
\end{table*}

\indent For different values of $\beta$ and $\xi$, $\Omega_{gw}$ of Gamma-HFGWs have
various distinctive envelopes with respect to rotational phase $\varphi$ (Fig.\ref{compareWaveEnvelopes}), but unlike the pulsars,
above envelopes will usually (but not always) come with two peaks
during every rotational period [for dipole-mode surface magnetic fields,
see Fig.\ref{compareWaveEnvelopes}(a), i.e. the
frequency of pulses is double with respect to the rotational frequency], or usually with four peaks for every rotational period [for
case with quadrupole-mode surface magnetic fields, see Fig.\ref{compareWaveEnvelopes}(b)], due to special angular
distributions of   Gamma-HFGWs (see Fig.\ref{compareWaveEnvelopes}).
Based on Eqs. (\ref{gammappfenvelopeBipole}) and
(\ref{gammappfenvelopeQuadrupole}), typical curves of $\Omega_{gw}$ envelopes of Gamma-HFGWs can be estimated
[see Fig.\ref{hfgwGammaSignalweak}(1)-(4) for  equator-maximum case and (5)-(8) for quadrupole case].
The characteristic envelopes would be helpful  to distinguish the Gamma-HFGWs
out of background noise, no matter for the model of magnetar magnetic fields we take here, or for other models with different structures.\\

\indent Given that the mechanism of GRBs is actually quite complex, more detailed issues of generation of HFGWs based on fireball model or other models such as Poynting flux
model \citep{Uso1992,Usov15041994,Smolsky1996,0004-637X-531-2-764,Drenkhahn2002,RevModPhys.76.1143},
involving magnetars or black holes or even other sources,  can be further studied as consequent future research projects.
However, we are providing some approximate estimations which can be addressed here for some situations. E.g., in various magnetar  and GRB  models, we always can  assume that effective (for HFGW generation)
EM radiations
decay by $r^{\Lambda}$ ($r$ is distance), and assume strong magnetic fields (contributing to HFGW generation) decay in $r^{\Xi}$. For typical $r^{\Lambda}$, $r^{\Xi}$ and
effective accumulation distance $D$ (for interaction between EM radiations and strong magnetic fields),
accumulated ``effective-Gamma-HFGWs'' generally have the form (for cases of $\Lambda +\Xi\neq2$):
\begin{eqnarray}
	\label{AeffectiveGammaHFGW}
	&~&A^{\gamma-eff}_{accum}(z)=\int^{D}_{r_0}\frac{8 \pi  G}{{kc}^5 {\mu_0}}\tilde{E}^{(1)burst}_{x.r_0}(\frac{r_0}{r})^{\Lambda} \nonumber\\
	&\cdot&B^{(0)surf}_{\theta-Max}(\frac{r_0}{r})^{\Xi}\cdot\frac{r}{z}dr \nonumber\\
	&=& \frac{8\pi G \tilde{E}^{(1)burst}_{x.r_0}B^{(0)surf}_{\theta-Max}}{kc^5\mu_0}\frac{1}{D}\frac{r_0^2-D^2(\frac{r_0}{D})^{\Lambda+\Xi}}{\Lambda+\Xi-2} ;
\end{eqnarray}
or for cases of $\Lambda +\Xi=2$, it is:
\begin{eqnarray}
	\label{AeffectiveGammaHFGW-2}
	&~&A^{\gamma-eff}_{accum}(z)= \frac{8\pi G \tilde{E}^{(1)burst}_{x.r_0}B^{(0)surf}_{\theta-Max}}{kc^5\mu_0}\frac{r_0^{\Lambda+\Xi}}{D}ln\frac{D}{r_0};\nonumber\\
\end{eqnarray}

Table \ref{effectiveGammaHFGWsTable} gives  estimations of energy density of above effective-Gamma-HFGWs at a given far field observation point, with short accumulation distance $D$ around the source of magnetar, given different effective
parameters $r^{\Lambda}$ and $r^{\Xi}$. Here, we are ignoring sources of Gamma-HFGWs outside the accumulation distance $D$. We find some of these energy density (around $10^{-6}$) would also suitable for the proposed HFGW detector \citep{FYLi_PRD67_2003,FYLi_EPJC_2008,LiNPB2016,FYLi_PRD80_2009,Li.Fang-Yu.120402}. \\

\section{Summary and discussion}
\label{sectSummary}

As powerful astrophysical bodies, magnetars may provide physical conditions leading to extremely strong celestial EM sources of HFGWs. This article attempts to address novel issues of generation of HFGWs (with very high-frequency$\sim10^{20}Hz$)
caused by interaction between ultra-high magnetar surface magnetic fields and strong radiations of GRBs.
We summarize the main results as follows:\\ 

\indent(1) We estimate the energy density $\Omega_{gw}$ of Gamma-HFGWs, and find that
for certain parameters of observational distance and GRB power, the Gamma-HFGWs would have far field $\Omega_{gw}$ around $10^{-6}$ (Table  \ref{gammaburstSignal}).  Gamma-HFGWs with such energy density could cause first-order perturbed signal EM waves of $\sim10^{-20}W/m^2$ in the proposed HFGW detection system based on EM response to
HFGWs and synchro-resonance effect \citep{LiNPB2016,FYLi_PRD67_2003,FYLi_EPJC_2008,FYLi_PRD80_2009,PRD104025,WenEPJC2014,Li.Fang-Yu.120402}. However, the issues arising as to how to extract and distinguish such perturbed EM signals from noise, and relevant concrete experimental techniques, are not key points in this paper, and they can be addressed in subsequent research studies and future works. At least, with studies of the far-field effect, we think that Gamma-HFGWs would provide possible potential targets of HFGWs for observation in the future from the Earth or from far field observation points.\\

\indent(2) More general and approximated estimations of generation of HFGWs by
GRB radiation
interacting with strong surface magnetic fields of a magnetar
have also been addressed. Brief derived estimations show that even if such general EM sources decay very fast (Table \ref{effectiveGammaHFGWsTable}),
they would still possibly lead to  $\Omega_{gw}\sim10^{-5}$ to $10^{-7}$ of HFGWs at an observational distance of $\sim3.3Mpc$, given typical effective accumulation distance and various decay ratios of the radiation and magnetic fields. Such levels would also be suitable for the proposed HFGW detector \citep{LiNPB2016,FYLi_PRD67_2003,FYLi_EPJC_2008,FYLi_PRD80_2009,Li.Fang-Yu.120402,PRD104025,WenEPJC2014}. \\

\indent(3) We find the envelopes of energy density of Gamma-HFGWs strongly depend upon the structure of surface magnetic fields of magnetars. E.g., for the model of magnetic fields of a magnetar we employ here (in dipole or quadrupole modes), the envelopes would appear in distinctive pulse-like patterns (see Figs.\ref{compareWaveEnvelopes}, \ref{hfgwGammaSignalweak}, based on estimated expressions of Eq. (\ref{gammappfenvelopeBipole}) and Eq. (\ref{gammappfenvelopeQuadrupole})).  In other words, such characteristic envelopes not only could deliver and reflect specific geometrical information of surface magnetic fields of the magnetars, 
but could also be an exclusive identification criterion to distinguish Gamma-HFGWs from background noise.\\

\indent (4) For the first step,  in this work we 
simply assume that the GRBs from magnetars radiate isotropically, so more specific angular distributions and physical
processes of GRBs should be adopted in the next steps. This  might also cause different strengths and envelopes of the Gamma-HFGWs. Besides, here we only focus on the dipole and quadrupole modes of the magnetar surface magnetic fields. In fact, several other models have also been proposed with different configurations of magnetar magnetosphere, e.g., some of them suggest twisted dipole \citep{APJ.574.332} instead of a centred dipole, 
or higher multipole components \citep{Pavan11052009}, or even more complicated structures
\citep{Zane01032006,Ruderman.APJ.1991}. Therefore, related
works concerning diverse patterns of HFGWs based on alternative models of magnetars or GRBs, would also be interesting topics for possible subsequent studies.\\
\indent If GRBs with different specific distributions are taken into account, the power of produced Gamma-HFGWs could decrease (if directions of  GRBs radiation and poloidal magnetic field do not match), or  could even increase (if GRBs are more concentrated in the direction perpendicular to the magnetic field, leading to more effective interaction).  Such variation and related models need to be verified by experimental observations. Nevertheless, our estimated results may sit in the sensitivity range of the proposed HFGW detector \citep{FYLi_PRD67_2003,FYLi_EPJC_2008,FYLi_PRD80_2009,PRD104025,LiNPB2016}, and could still allow some room for considering a more relaxed parameter range and some alternative models.  However, experimental issues are not the key point of this study, and detailed research for such issues should be carried out later.\\

\indent In general,  magnetars could be involved in possible astrophysical EM sources of GWs in very high-frequency bands, and the Gamma-HFGWs they produce would provide far field effects with distinctive characteristics, so they would be possible potential targets for observation in the future.
If any Gamma-HFGWs can be detected, they may provide evidence not only for HFGWs from super powerful astrophysical process and celestial bodies,  but also provide us with astrophysical benchmarks which we can use as references
for different models of magnetars (including their inner structures and configuration of surface magnetic fields). We  anticipate future research work and development of additional models of GRBs for future gravitational wave astronomy investigative work. \\

\begin{acknowledgments}
This work is supported by the National Natural Science Foundation of China (No.11605015, No.11375279, No.11205254, No.11647307), and the Fundamental Research Funds for the Central Universities (No.106112017CDJXY300003 and 106112017CDJXFLX0014).
\end{acknowledgments}

\bibliographystyle{apsrev4-1}
\bibliography{WenReferenceData20150101}
\end{document}